\documentstyle[12pt]{article}
\title{Horava-Lifshitz gravity, absolute time, and objective particles in curved space}
\author{Hrvoje Nikoli\'c \\
Theoretical Physics Division, Rudjer Bo\v{s}kovi\'{c} Institute, \\
P.O.B. 180, HR-10002 Zagreb, Croatia \\
{\normalsize hrvoje@thphys.irb.hr} \\
\makebox[1in]{} \\
}
\date{\today}
\begin{document}
\maketitle
\begin{abstract}
Recently, Horava formulated a renormalizable theory of quantum gravity that 
reduces to general relativity at large distances but violates
Lorentz invariance at small distances. The absolute time involved in this theory
allows to define an objective notion of particles associated with 
quantization of fields in classical gravitational backgrounds. The Unruh effect
and other observer-dependent notions of particles in curved space are interpreted as effects
caused by interaction between the objective vacuum and
the measuring apparatus made up of objective particles.
\end{abstract}


\section{Introduction}

Recently, Horava \cite{horava} (influenced by old ideas of Lifshitz \cite{lifs} on 
phase transitions in condensed-matter physics)
formulated a new field theory of gravity that 
reduces to general relativity at large distances. However, 
at small distances this theory violates Lorentz invariance, in such a way that the corresponding
quantum theory of gravity becomes renormalizable.
Owing to these remarkable properties, this Horava-Lifshitz
theory of gravity is receiving a considerable attention 
\cite{razno1}-\cite{keh2},
both in the quantum and classical regime.

In this paper, we discuss some implications of the Horava-Lifshitz theory on semi-classical
gravity, i.e., on the regime in which
matter fields are quantized while gravity is treated as a classical background.
More precisely, we discuss how the absence of the Lorentz invariance in
Horava-Lifshitz theory helps to solve the old 
problem \cite{full} of defining particles in classical gravitational backgrounds.
(For an explicit calculation of Hawking radiation in Horava-Lifshitz gravity see \cite{chen}.)
The next section is a brief overview of the problem of particles in curved spacetime
from the standard point of view, while the discussion of the same problem from the point of view of Horava-Lifshitz theory is presented in Sec.~\ref{SEC3}.

\section{The problem of particles in curved spacetime}

The problem can be formulated as follows \cite{bd}. A scalar field operator
$\phi(x)$ satisfying the Klein-Gordon equation in curved spacetime
(with the signature $(-,+,+,+)$)
\begin{equation}
 [\nabla^{\mu}\nabla_{\mu}-m^2]\phi(x)=0
\end{equation}
can be expanded as
\begin{equation}
 \phi(x)=\sum_k a_k f_k(x) + a_k^{\dagger} f^*_k(x) ,
\end{equation}
where $\{f_k(x), f^*_k(x) \}$ constitute a complete and orthogonal basis
of the classical solutions of the Klein-Gordon equation, while $a_k, a_k^{\dagger}$ are
operators satisfying the commutation relations
\begin{equation}\label{comrel}
 [a_k, a_{k'}]=[a^{\dagger}_k, a^{\dagger}_{k'}] =0 , \;\;\;\;
[a_k, a^{\dagger}_{k'}]=\delta_{k,k'} .
\end{equation}
The commutation relations (\ref{comrel}) suggest the particle interpretation,
according to which the vacuum $|0\rangle$ satisfies 
\begin{equation}
 a_k|0\rangle =0,
\end{equation}
while $n$-particle states are defined by acting $n$ times with the creation operators
$a^{\dagger}_{k}$ on the vacuum. The problem is that the basis $\{f_k(x), f^*_k(x) \}$
can be introduced in many different ways, which leads to many different 
definitions of particles. The different definitions of particles are not equivalent, in the sense
that the number of particles in a given quantum state depends on this definition.
In particular, the vacuum defined with respect to one definition of particles may be
a many-particle state with respect to another definition of particles \cite{bd}.
In Minkowski spacetime the particles can be defined with respect to the basis in which
the functions $f_k(x)$ have a positive frequency. However, the notion of 
frequency depends on the choice of the time coordinate. On the other hand, the principle
of covariance with respect to arbitrary spacetime coordinate transformations, which plays
a crucial role in general relativity, implies that there is no any natural choice of the
time coordinate. Consequently, there is no any natural definition of particles either.

To overcome this problem, one possibility is to adopt the point of view according to which
particles do not play any fundamental role in quantum field theory in curved spacetime
\cite{bd,wald}. However, such a point of view does not seem to be completely satisfying
because almost everything that we observe in high energy experiments seems 
to be defined in terms of particles. 
Another possibility is to interpret the concept of particle
as an observer dependent entity, because each observer may have his own natural definition
of time. However, such a definition of particles
seems to be somewhat vague because the concept of an observer is not well defined in
physical terms. Moreover, in a transition from one observer to another the concept of particles 
does not transform in a covariant manner. Consequently, as stressed in \cite{unruh},
it is not clear which definition of particles,
if any, corresponds to the particles that contribute to the covariantly transforming
energy-momentum tensor that determines the gravitational field.

\section{Particles in Horava-Lifshitz gravitational background}
\label{SEC3}

Now, from the point of view of Horava-Lifshitz theory of gravity, the problem of particles
takes a completely different flavor. The Horava-Lifshitz theory of gravity
is not covariant with respect to arbitrary spacetime coordinate transformations.
The fundamental fields that describe gravity are the spacial metric 
$g_{ij}({\bf x},t)$, the lapse function $N({\bf x},t)$, and the shift vector
 $N_i({\bf x},t)$, where $i,j=1,2,3$ and ${\bf x}=\{x^1,x^2,x^3\}$.
Here we do not need the explicit form of the action $S[g_{ij},N,N_i]$
because we treat
$g_{ij}({\bf x},t)$, $N({\bf x},t)$, $N_i({\bf x},t)$ as a fixed non-dynamical background.
It suffices to say that the action \cite{horava} is covariant with respect to 
coordinate transformations that do {\em not} mix space coordinates $x^i$
with the time coordinate $t$. However, the action is not covariant with respect to 
coordinate transformations that do mix space and time coordinates. 
Such a time-space mixing covariance appears only in the  
low energy limit. The fundamental fields $N$, $N_i$ define the components
of an effective spacetime metric through the Arnowitt-Deser-Misner relations
\begin{equation}
N=(-g^{00})^{-1/2}, \;\;\;\; 
N_i=g_{0i},
\end{equation}
but due to the lack of general spacetime coordinate invariance, the notion of spacetime
does not longer play any fundamental role. Instead, space and time are 
fundamentally independent entities. In particular, time is absolute (up to a 
trivial rescaling of the form $t'=f(t)$).
By having an absolute notion of time, one can also define an objective notion of
particles, as particles defined with respect to functions $f_k({\bf x},t)$ which have a positive
frequency with respect to the absolute time $t$.

Even though the particles in quantum field theory now have an objective status,
it does not mean that the number of particles cannot change with time.
Indeed, when the gravitational background is time dependent, one expects particle 
creation \cite{bd}. The particle creation can be calculated in an exactly the same way
as in the conventional approach with a fixed gravitational background \cite{bd}, 
provided that the correct time coordinate has been identified. For example, the standard
derivation of Hawking radiation \cite{bd,hawk} is based on the assumption 
that the particles outside the horizon are defined with respect to the Schwarzschild time.
If this identification of time is the correct one from the point of view of Horava-Lifshitz theory, then these Hawking particles are objective.
Indeed, as shown in \cite{keh1,keh2}, the classical spherically-symmetric solution  
\begin{equation}\label{metric1}
 ds^2=-N^2(r)dt^2 + \frac{dr^2}{f(r)}+r^2(d\theta^2+\sin^2\theta \, d\varphi^2)
\end{equation}
of Horava-Lifshitz theory in the limit of vanishing cosmological constant reads
\begin{equation}\label{metric2}
 N^2(r)=f(r)=1+\omega r^2 - \sqrt{r(\omega^2r^3+4\omega M)} ,
\end{equation}
where $\omega$ is a free parameter of the theory. For large $r$ it reduces to
\begin{equation}
f(r) \approx 1-\frac{2M}{r} +{\cal O}(r^{-4}) ,
\end{equation}
which corresponds to the Schwarzschild metric. The exact metric (\ref{metric2})
contains two horizons at
\begin{equation}
 r_{\pm}=M\left( 1\pm \sqrt{1-\frac{1}{2\omega M^2}} \right) .
\end{equation}
Assuming that $M^2 \gg \omega^{-1}$, the outer horizon $r_+\approx 2M$
is close to the Schwarzschild horizon, while the inner horizon  $r_-\approx 0$
is close to the singularity. 
Since we are not allowed to transform (\ref{metric1}) to other spacetime 
coordinates by coordinate transformations that mix $r$ and $t$, 
this shows that the Schwarzschild time is indeed the preferred time
of the black hole and that the singularity on the horizon is not merely 
a coordinate singularity.
This means that, contrary to the usual interpretation of Hawking radiation, all observers 
should agree that the black hole radiates, including the freely falling observers near the horizon.
(We also point out that even the notion of an event horizon does not play the usual 
role in Horava-Lifshitz gravity \cite{horava} because the absence of Lorentz invariance
implies that information can travel even faster than light, allowing information to escape
from the horizon. However, it does not seem to affect the creation of Hawking radiation, 
because only the existence of an apparent horizon is essential for Hawking mechanism  
of particle creation \cite{viss}.)
We also note that with an absolute notion of time, the particles in a gravitational background
can also be defined locally in terms of local particle currents that are not conserved at the points
at which the particle creation takes place \cite{nikpart1,nikpart2,nikpart3}. 

Now, if particles in a gravitational background are objective entities, then how to interpret 
the Unruh effect \cite{unruh,bd}?
The Unruh effect is a statement that the Minkowski vacuum appears as a thermal state with an
indefinite number of particles when viewed by a uniformly accelerated observer.
There are two approaches how the Unruh effect can be derived \cite{bd}. The first approach is 
to define the particles with respect to time of an accelerated observer, leading to the notion
of Rindler particles. The second approach is to define the particles with respect to the Minkowski time,
but to study a response of an accelerated detector. 
The Minkowski spacetime is a solution of Horava-Lifshitz theory,
which corresponds to $M=0$ in (\ref{metric2}). Again, we are not allowed
to transform this solution to Rindler coordinates, because such a coordinate transformation
would mix time and space coordinates. 
Consequently, the objective particles in a flat Horava-Lifshitz background are the Minkowski particles 
so only the second approach above is the correct one. Indeed, it has been
already shown that the two approaches are not equivalent \cite{rind1,rind2}, so it is impossible that both approaches are correct.

The Unruh effect and its generalizations in various gravitational backgrounds can be more generally
understood as follows. The vacuum (i.e., the state without objective particles) can be
expanded as 
\begin{equation}\label{u3.3}
 |0\rangle = \sum_{n=0}^{\infty}\sum_{\xi_n} c_{n,\xi_n} |n,\xi_n\rangle'
\end{equation}
where $c_{n,\xi_n}$ are some coefficients of the expansion and 
$|n,\xi_n\rangle'$ are ``$n$-particle'' states defined with respect to some
other time coordinate corresponding to a spacetime coordinate transformation
with respect to which the Horava-Lifshitz gravity is not covariant.
The parameter $\xi_n$ parameterizes different states with the same 
``number of particles'' $n$. For $n=0$, $\xi_n$ takes only one possible value,
i.e., $|0,\xi_0\rangle'\equiv |0\rangle'$.
The quantity $|c_{n,\xi_n}|^2$  is the probability that the system will be found in the state $|n,\xi_n\rangle'$, but it does not have any operational meaning without having a 
physical interpretation of $|n,\xi_n\rangle'$. In fact, the state (\ref{u3.3}) 
receives an operational meaning only if it becomes entangled with a macroscopic
measuring apparatus, or more generally, with a macroscopic environment.
If the total state describing the entanglement with the environment takes the form 
\begin{equation}\label{u3.4}
 |\Psi\rangle_{\rm total} = \sum_{n=0}^{\infty}\sum_{\xi_n} c_{n,\xi_n} 
|n,\xi_n\rangle' |E_{n,\xi_n}\rangle ,
\end{equation}
where $|E_{n,\xi_n}\rangle$ are macroscopically distinct states of the environment
(measuring apparatus) describing a large number of objective 
particles, then $|c_{n,\xi_n}|^2$ is the probability that the environment
will be found in the state $|E_{n,\xi_n}\rangle$. 
If the environment is found in the state $|E_{n,\xi_n}\rangle$, then 
it is justified to say that the measured system described by (\ref{u3.3}) is found
in the state $|n,\xi_n\rangle'$.
When the environment is accelerated or when a gravitational background is present, 
the theory of decoherence can explain why 
the interaction with the environment leads to an entanglement of the 
form of (\ref{u3.4}) \cite{decohunruh1,decohunruh2,decohunruh3}, with $|n,\xi_n\rangle'$ being the 
``$n$-particle'' states defined with respect to the proper time of the environment.
In particular, if the environment is found in the state $|E_{0}\rangle$, then 
the measured system is in the state $|0\rangle'$. This state is actually 
a squeezed state \cite{grish}, containing an uncertain number
of objective particles with an average number of objective particles being larger than zero.
Thus, the interaction with the environment creates new objective particles,
even when the measurement apparatus is found in the state $|E_0\rangle$.

To conclude, we have seen that the Horava-Lifshitz theory of gravity
significantly reinterprets the old problem of particles associated 
with quantum fields in gravitational backgrounds. 
The absolute time of the Horava-Lifshitz gravity suggests that particles
in quantum field theory retain their objective status, similarly to the more familiar
concept of particles associated with quantum fields in Minkowski spacetime.

\section*{Acknowledgements}
This work was supported by the Ministry of Science of the
Republic of Croatia under Contract No.~098-0982930-2864.

\end{document}